\documentclass[final,5p,twocolumn,times]{elsarticle}
\usepackage[T1]{fontenc}
\usepackage[utf8]{inputenc}
\usepackage{graphicx}
\usepackage{amssymb}

\usepackage{amsthm}

\newcounter{bla}

\journal{Computer Physics Communications}

\begin{document}

\begin{frontmatter}



%\title{PASYVAT: Topological and geometric structure characterization of particle systems through selective multiple range bonding, using radial distribution function}

\title{Visual, user-interactive generation of bond networks in 3D particle configurations}

\author[llnl]{Alfredo Metere\corref{cor1}\fnref{fn1}}
\ead{metere1@llnl.gov}
\fntext[fn1]{Telephone: +1 (925) 423-6913}

\author[llnl]{Tomas Oppelstrup}
\author[nada]{Mikhail Dzugutov}
\author[uppsala]{Anders Hast}
%\ead{mik@csc.kth.se}

\cortext[cor1]{Corresponding author.}

\address[llnl]{Lawrence Livermore National Laboratory, 7000 East. Ave., Livermore, CA 94550-9234, USA}

\address[nada]{Department of Mathematics, Royal Institute of Technology (KTH), 100-44 Stockholm, Sweden}

\address[uppsala]{Uppsala University, Lägerhyddsvägen 2 - Hus 2, 751 05 Uppsala, Sweden}

\hyphenation{PASYVAT}

\begin{abstract}
We present a new program able to perform visual structural analysis on 3D particle systems called PASYVAT (PArticle SYstem Visual Analysis Tool). More specifically, it can select multiple interparticle distance ranges from a radial distribution function (RDF) plot and display them in 3D as bonds between the particles falling within the selected distance range, thus generating a network of bonds. This software can be used with any data set representing a system of points or other objects having a well-defined center of mass or geometric center in 3D space. In this article we describe the program and its internal structure, with emphasis on its applicability in the study of certain particle configurations, obtained from classical molecular dynamics simulation in condensed matter physics.
\end{abstract}

\end{frontmatter}

\section{Introduction}
\label{introduction}
The visual analysis of a structure is an increasingly important practice in the characterization of results coming from computer simulations. The analysis of simulation data is often achieved through numerical command-line programs. For this reason they often require an extensive prior understanding of the expected results. Furthermore, numerical analysis can be time inefficient, if the purpose of the characterization is to obtain a preliminary, qualitative overview of the obtained result. Such techniques, known as analytical (or numerical) methods, are common in diverse areas of science, including astrophysics, condensed matter physics, materials science and materials chemistry.
By directly monitoring the phase-space evolution of a system of interacting particles, the simulation provides insight into the system's dynamical processes and structural evolution which cannot be obtained from the laboratory experiments. In condensed-matter physics and materials science, particle simulations address increasingly complex and realistic problems, gaining ground as an indispensable ingredient of materials research and development.
A simulation of a particle system produces a set of coordinate files representing system's configurational evolution. A central problem in the analysis of these data is to reduce the vast amount of information contained in the coordinate files to a comprehensible pattern that would make it possible to identify the structure. For that purpose, the use of graphical software is indispensable for generating visual patterns of interparticle bonds that allows the recognition of the structure in terms of bonding topology.\\

Calculation of radial distribution function (RDF), $g(r)$, is the first and the simplest step in the structure analysis of a simulated particle system. $g(r)$ can be viewed as the normalized probability density that two particles are separated by a distance $r$. Then the ensemble averaged number of neighbors of a given particle confined within the distance $r$ can be expressed as:   

\begin{equation}
n(r) = 4 \pi \rho  \int_0^r x^2 g(x) dx
\end{equation}
\label{eq_avgens}

where $\rho=N/V$ is the average number density for a system of $N$ particles occupying volume $V$. Using this expression, an operational definition of RDF through $n(r)$ can be inferred:

\begin{equation}
4 \pi r^2 \rho g(r) = d n(r)/ dr
\end{equation}
\label{eq_rdfop}

Once RDF is calculated, the information it contains can be used to select the range of interparticle distances for generating the bond network.\\

PASYVAT is a software package designed for topological and geometric structure characterization of particle systems, based on visually analyzing the data set of interest. It is based on the selection of multiple bonding ranges; this selection is done based on the visual analysis of RDF plot.
The relevant aspects of the topology pattern can be selectively analyzed by using an appropriate range of bond lengths from the RDF plot. 
First, a preliminary, qualitative inspection of the RDF plot features is necessary for identifying the relevant ranges of bond lengths. Then, interparticle distance boundaries confining those features can be selected, respectively.
This method can find application in the topological analysis of systems having different physical nature, ranging from the microscopic order in condensed matter systems to aggregations of astronomical objects.\\

By analyzing the topologies of the networks of bonds generated for different ranges of interparticle distances, we can classify the real structure observed in materials and physical systems. The unique advantage of using PASYVAT resides in the direct linking of the RDF features to the generated bond network. This makes it possible to visually recognize the underlying archetypal structure and identify the topology class to which the investigated system structure belongs.
The topological classification of structures by their bond network is an important characterization method for porous materials. There is a whole new branch of chemistry that emerged from this approach, called \textit{reticular chemistry}. Its industrial applications are of potentially vast importance, including the synthesis of novel porous materials for hydrogen storage or carbon dioxide separation. The study of isoreticular structures and tailored porous structures is developing fast and thousands of topological patterns, found in both real and hypothetical structures, have been published \cite{rcsr_database}.
One of the main ideas considered for the design of PASYVAT was to make the program easy to use and modify. For this reason the program allows only operations that can be done through Graphical User Interface (GUI), without the need of writing scripts or command-line input. Another important feature of this software is universality. The program does not a priori assume any structural order in the particle configuration to be analyzed. For this reason the PASYVAT program is quite different from some crystallography software packages that have been used in the past for the topological characterization of the particle configurations produced by molecular dynamics simulation \cite{misha}. In fact, crystallography programs usually assume in advance that the investigated configuration is a chemical lattice, with specified symmetry and atomic parameters. If the particle system was not produced with crystallographic methods and only the position in space is defined for each particle, to guess the topological features becomes a non-trivial operation. A significant example is represented by those particle systems coming from the molecular dynamics simulations of simple liquids\cite{misha1}. 
A known limitation of most crystallographic programs, when used on generic particle systems, consists in the restrictive maximum number of particles that the program can load and process at once. This restriction partially depends on the structure of the software, which has not been designed for large particle configurations, like the ones typically used in molecular dynamics. PASYVAT has been thought to be generally applicable to any data set representing a system of particles. Hence, regardless of their physical meaning, the particles in the input data structure are treated as simple points individuated only by a three-component vector representing their position in a 3D Euclidean space.\\

In an effort to keep PASYVAT general and independent from the size of the data set, 
the software has no imposed restrictions either in the input file size or in the precision of the numerical calculations used for the structural analysis.
With the same purpose in mind, PASYVAT has been implemented using the Visualization Tool Kit (VTK) \cite{vtk} libraries, which support MPI parallel rendering. Despite VTK libraries present the drawback of being mostly implemented with the now obsolete fixed-pipeline rendering, they are rich in powerful 3D analysis tools. This was deemed, to the best of our knowledge, as an acceptable compromise for most of the real-case scenarios. \\

\section{Program Overview}
\label{p_overview}
\begin{figure*}
  \centering
   \setlength\fboxsep{0pt}
   \setlength\fboxrule{0pt}
    \fbox{\includegraphics[width=6.25in]{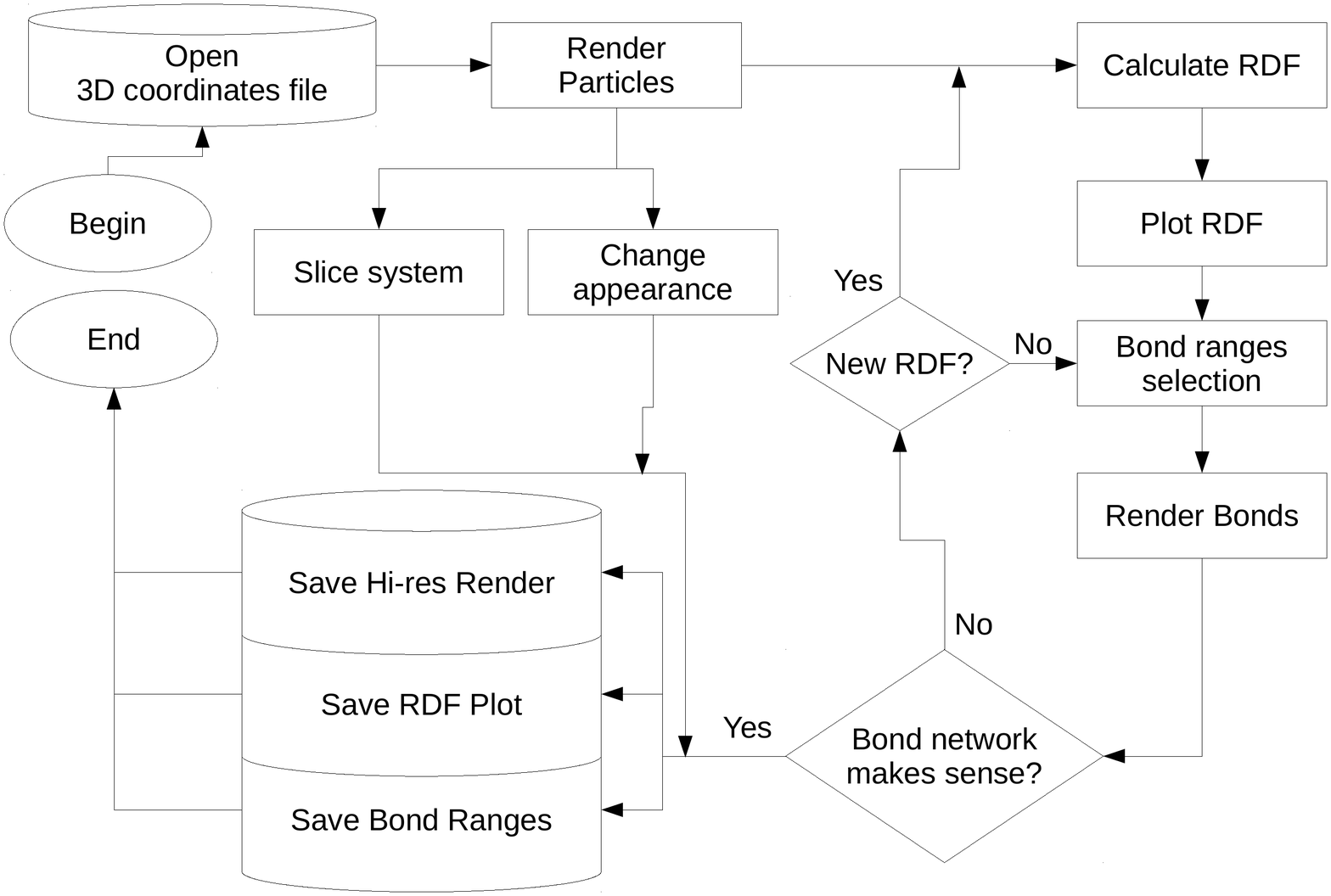}}
    \caption{A flowchart representation of the modules composing PASYVAT.}
    \label{fig1}
\end{figure*}

The central element of PASYVAT consists of its GUI. Through the GUI it is possible to open an input file containing particle configuration, visualize it as a real-time rendered 3D representation and immediately calculate its RDF.
The user is allowed to tune the RDF calculation parameters, to identify the bond connectivity between the particles of the system. Each bonding pattern is determined for an assumed range of interparticle distances, which is chosen to include an arbitrarily selected feature of RDF. This procedure is user controlled and can be repeated in order to get a comprehensive general picture of bond topology in the system. The general picture of the bond connectivity topology in the investigated system of particles, obtained in this way, provides an important insight in the general structural characterization of the system. Fig. \ref{fig1} shows how the main modules of PASYVAT are interconnected.

\section{Program Workflow}
\label{p_workflow}

Once the program is launched, it shows the main window in the user screen. From the main window menu, the user can open a file containing the 3D coordinates of the configuration and visualize the particles as a set of spheres with fixed radius. \\

The program checks the integrity of the coordinate file and calculates the dimensions of the parallelepiped which contains the particle configuration. The user can interact with the render window to rotate, zoom and translate the particle system. After the particle system is loaded, the user can calculate its RDF, choosing the maximum radius and the RDF plot resolution. \\

The parameters used in the calculation of the RDF are then reported in the terminal window and a CSV file containing the RDF plot coordinates is saved on the disk. The user can choose to visualize the calculated RDF as interactive plot in a separate window. From the window containing the RDF interactive plot, the user can select and adjust the boundaries of two separate ranges of interparticle distance. Once the RDF features of interest are selected in the plot window, the user can request PASYVAT to draw segments between the particles situated at a distance within those selected ranges.
The lines appearing in the 3D real-time rendered window will likely form a certain interparticle connectivity pattern or  network.
For this reason, they will be referred from now on as ``bonds''.
A unique bond color is associated to each of the interparticle distance ranges selected (usually two).\\

After the bond generation, to facilitate the detection of interesting geometries formed by the selected bond pattern, the user can interact with the render window by rotating, scaling and translating the bond network model. When the user is satisfied with the results produced, it is possible to save a high-resolution, press-quality snapshot of the render window. The image file generated has dimensions fixed to ten times the size in pixels of the render window viewport.\\

It is also possible to save the RDF plot as Portable Document Format (PDF) file, as well as Portable Network Graphics (PNG) image file or simply print it. The boundary values for each of the bond ranges can be saved in a text file.
These output formats enable the user to publish press-quality images of the results.

\section{Program Implementation}
\label{prog_imp}
The PASYVAT program has a modular structure, implemented in Python. To best accomplish this modular structure design, most of the source code has been developed following the Object-Oriented Paradigm (OOP), which is native for Python programming language. The numerical analysis subroutines are written in free-form Fortran 90 and wrapped to Python. Instead, these numerical subroutines are developed following the procedural programming paradigm, because Fortran is natively a procedural language. This means that implementing OOP code in Fortran, even if technically possible, it would result in no benefits, in the context of PASYVAT development. The interface between these procedural subroutines and the main program consists of wrapper libraries.\\

\subsection{Python for the core}
Python has been chosen as the main program language because it is well supported, well maintained, Open-Source, modern, extensible, powerful, flexible and portable. Python is a high-level, object-oriented, interpreted, general purpose programming language, featuring software interfaces to system calls, libraries and various windowing systems \cite{python}. These software interfaces give Python the possibility of take advantage of libraries which are significantly more performant than native Python calls, because they are often written in C, C++, or Fortran. The most common software interfaces used in Python are wrapper libraries and they are described more in detail at the subsection \ref{wrappers}.

\subsection{Fortran for numerics}
Fortran is a numerical, high-level, compiled language. It has been and still it is extremely popular among scientists because the syntax is clear and readable. Another important feature of Fortran is that the compilers can produce very efficient machine language code. Fortran can therefore perform much faster than native Python code. Free-form Fortran 90 is a very popular standard variant of Fortran and it has been chosen because of its improved readability and portability. The numerical analysis orientation of Fortran programming language makes it unsuitable for the development of programs with a complex user interface, like a GUI. This means that it is possible to implement programs like PASYVAT completely in Fortran, but it would take an unacceptable amount of time. Because of this, Fortran has been used to develop only the computationally intensive subroutines present in PASYVAT. These subroutines consist of the RDF calculation, the bond range selection and the bond indexing. These subroutines will be discussed more in detail later.

\subsection{Wrapper libraries}
\label{wrappers}
A wrapper library consists of a thin layer of code which translates the existing software interface of a library written in C, C++, Fortran or other programming languages into a Python-compatible software interface. This procedure in fact relieves the Python developer from understanding the higher complexity of those libraries written originally in C, C++, Fortran or other programming languages \cite{python}. Wrapper libraries can be generated by a code wrapping software utility, with different degrees of automation or completely manually. The procedural subroutines, written in Fortran, have been automatically wrapped into the Python code using the $f2py$ utility, which is part of the Numerical Python extensions (NumPy) \cite{numpy}. The utility $f2py$ scans the Fortran source and generates a signature file, containing all the relevant information needed to generate the wrapper code. Next, it generates the wrapper code, based on the signature file. The task is completed by the compilation of all the sources and the generation of a Python extension module containing the wrappers and the library \cite{f2py}.\\

\subsection{Graphical User Interface}
The GUI is implemented using the Qt application framework as a widget toolkit. There are many different widget toolkits and application frameworks for GUI implementation which are wrapped into Python. Qt has been chosen for being free, cross-platform and the because it features the $guiqwt$ toolkit \cite{guiqwt}, which has been extensively used in PASYVAT. Qt comes in different licensing and code variants: PASYVAT uses the free Open Source version with LGPLv3 license. The GUI is structured as a Multiple Documents Interface (MDI). The main window class instantiates the other GUI elements as public objects, like the render window class, the RDF plot window class and the RDF calculation dialog class.\\

\subsection{3D Rendering}
The Visualization Tool Kit (VTK), version 5.8.0, constitutes the foundation library for the 3D rendering engine used in PASYVAT. VTK is a collection of functions, classes, modules and interfaces dedicated to scientific visualization. The toolkit is entirely written in C++ and it is interfaced to Python via wrapper libraries. A relevant feature for the PASYVAT implementation is its compatibility with the main publicly available application frameworks, including Qt. This toolkit is oriented to scientific data visualization and it is currently developed and maintained by Kitware. It is distributed for free and it is open-source. VTK features parallel rendering for shared memory and distributed memory model computer architectures. The parallel rendering is implemented by combining the OpenGL calls with the standard Message Passing Interface (MPI). This means that PASYVAT inherits from VTK the capability to render in real-time large particle systems on High Performance Computing (HPC) class visualization centers. Currently, VTK's poor support for OpenGL Shading Language (GLSL) shaders makes it unsuitable for taking full advantage of the modern GPU hardware.\\

\section {Analysis Functions}

\subsection{Multiple interparticle distance ranges selection and drawing from RDF plot function}
The unique, novel function of PASYVAT is the interactive selection and drawing of multiple interparticle distance ranges from RDF plot. Once the particle system configuration is loaded in the program, the user interacts with the GUI to open the RDF calculation dialog. In the RDF calculation dialog it is possible to select the number of bins for the RDF and its distance range.
After the calculation of the RDF is completed, the plot button becomes enabled and when clicked the RDF plot window will appear. The user is free to simultaneously generate and open many RDF plots.\\
From the RDF plot window, it is possible to select the interparticle distance ranges, which appear highlighted in red. Through selecting the manipulation tool from the toolbox panel located right on top of the plot, the user can zoom and pan the RDF plot, as well as move and resize the selection ranges. \\
Once the user is satisfied with the selected ranges, clicking on the "Calculate Bonds" button will provoke PASYVAT to search for those interparticle distances falling within the highlighted range and plot the interparticle distances as cylindrical segments in the 3D window.\\
The whole operation requires a few seconds for 16384 particles and ~10000 bins (very high resolution) RDF in an average Linux desktop computer.\\

By default PASYVAT is set to allow the selection of only two distinct interparticle distance ranges; however the program can be easily adjusted to support as many as needed.

\begin{figure}
a) \includegraphics[width=8cm]{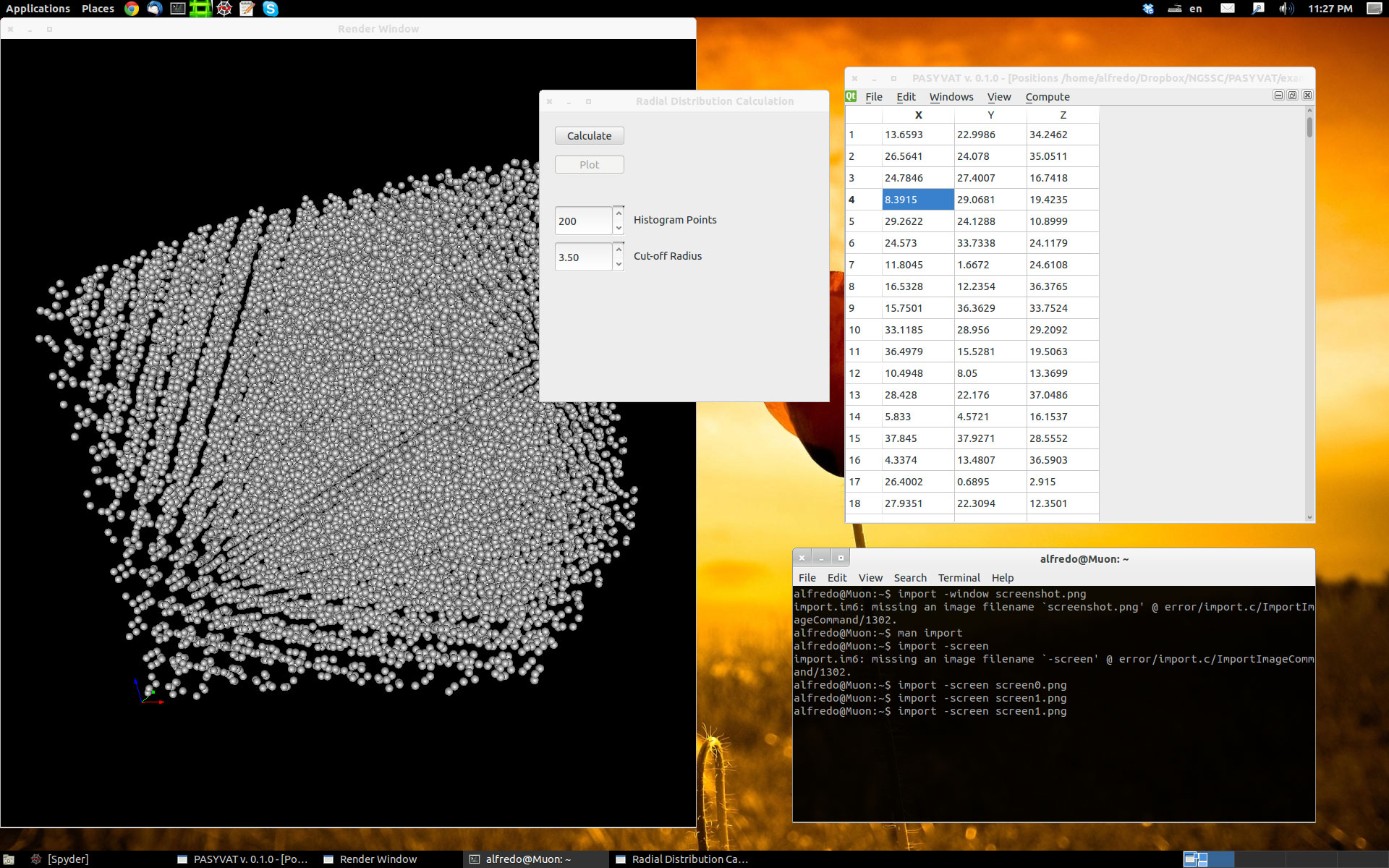}\\
b) \includegraphics[width=8cm]{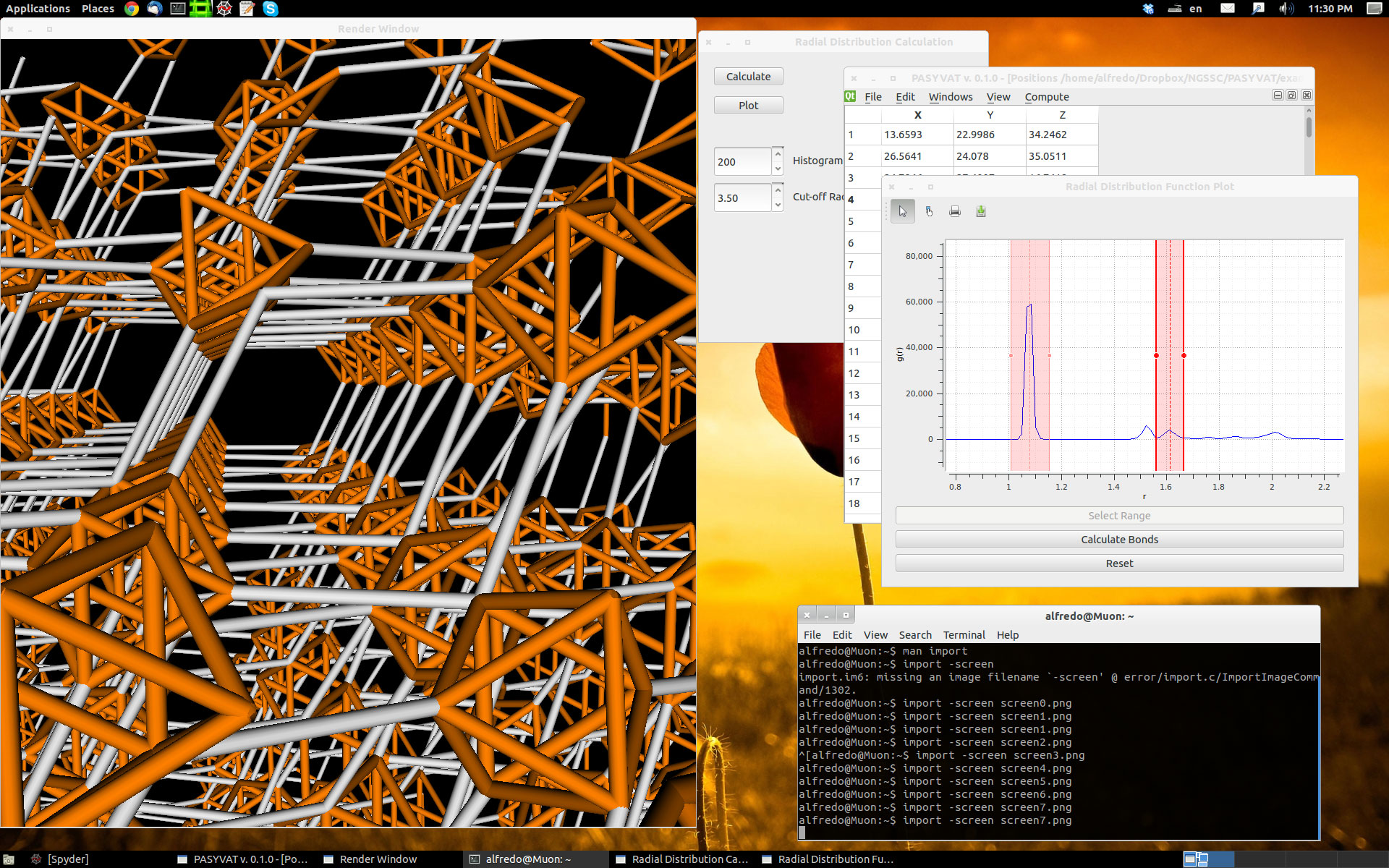}
\caption{Screenshots of PASYVAT. Example of RDF multiple interparticle distance range selection and drawing tool. $a)$ The raw simulation data. - $b)$ The bond network obtained from selecting multiple interparticle distance ranges from the RDF plot.}
\label{fig2}
\end{figure}

This function makes PASYVAT indispensable\cite{mofpaper} to successfully complete the characterization of non-trivial bond networks (Fig. \ref{fig2}).

\begin{figure}
a) \includegraphics[width=8cm]{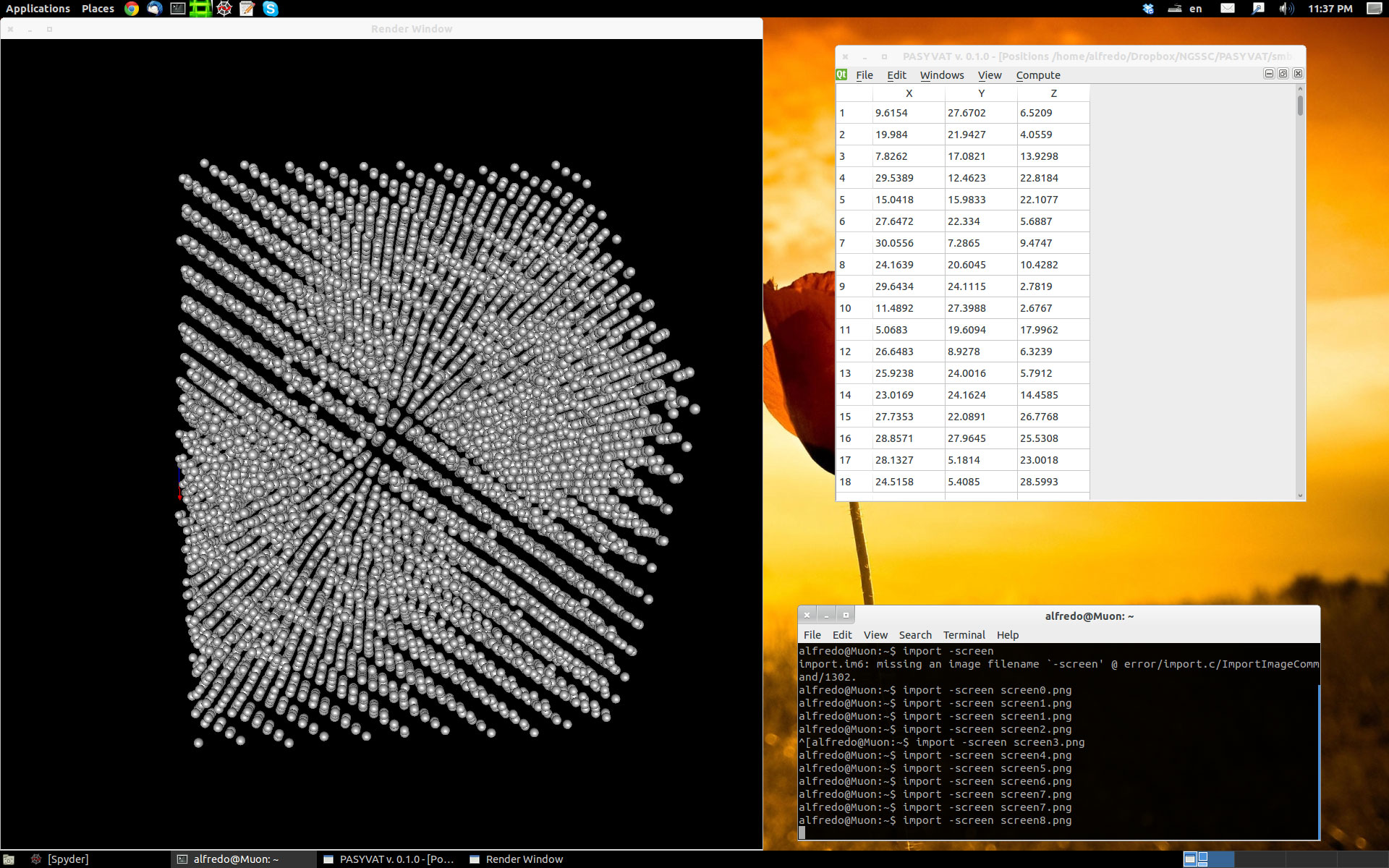}\\
b) \includegraphics[width=8cm]{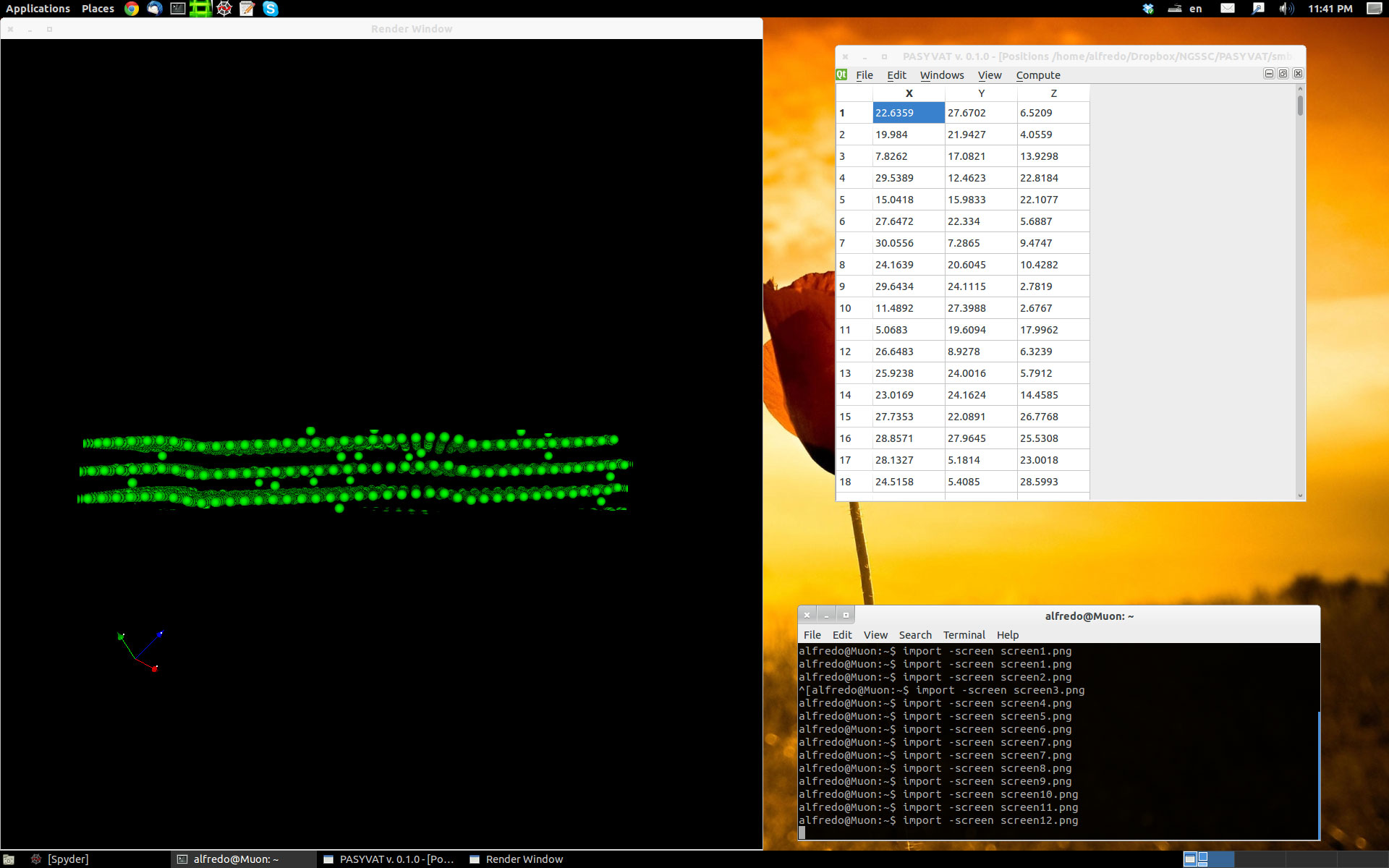}
\caption{Screenshots of PASYVAT. Example of slicing tool. $a)$ The raw simulation data. - $b)$ The sliced structure.}
\label{fig3}
\end{figure}

\subsection{Structure slicing function}
An useful analysis tool which has been implemented in PASYVAT is the structure slicing function (Fig. \ref{fig3}). Despite this tool lacks of novelty, it has been implemented in PASYVAT out of necessity \cite{smb}. This function can be toggled only via key-press events while the render window is focused. The tool draws a box enclosing the particle configuration. This box features four box-resize handles, which are represented as white spheres centered respect to the faces of the box and connected by lines appear. Left-clicking and dragging with the mouse on any of the box-resize handles will cause the box to change shape and volume, respectively. By clicking and dragging with the left mouse button on any other point of the box face will provoke a 3D rotation of the box around its center. All the particles that are left outside the faces of the box will be not rendered. 
Once the slice has been generated, the box and the box-resize handles could be toggled on or off anytime to improve the visibility of the sliced data set. 
The slicing function is based on the 3D Box Widget of VTK.\\

The performance is real-time when the slicing tool is used together with point representation of particles. Switching to polygonal low-resolution spheres will provoke a noticeable slowdown, but with 16384 particles and a very fast GPU is still possible to operate.\\

\section{Conclusions}
PASYVAT is effective for the analysis of certain particle configurations where particle connectivity is key to structure classification. It has been successfully used to analyze and characterize several particle configurations representing physical systems in which the chemical description was missing.
Such systems are commonly used in classical molecular dynamics simulation of self-assembled condensed matter phases \cite{smb} \cite{mofpaper}.\\

Despite being already useful, the program is still under active development for the implementation of both additional structure analysis functions and performance improvements of the current visualization engine.
A new 3D real-time visualization engine, entirely written from scratch in C++ will substitute VTK in the future releases. This choice has been dictated by the unacceptably poor support of GLSL in VTK. The new engine is already capable of rendering in real time 100 million particles on a single, average desktop computer.\\

Another scheduled improvement of PASYVAT consists of an MPI-parallel implementation for the new, high-performance visualization engine.
PASYVAT will also feature additional file I/O modules for expanding its compatibility with the existing simulation and visualization programs.
The RDF calculation routine is undergoing optimization and porting to CUDA.

\section{Where to find PASYVAT}
PASYVAT is Open Source and can be downloaded free of charge, under LGPLv3 licensing, from a GitHub repository at the following link:\\

http://www.github.com/alfredometere/PASYVAT/

\section{Acknowledgments}
The corresponding author gratefully acknowledges Prof. Hildred Crill for the support in revising the language. This work has been approved for release under Lawrence Livermore Release No. \#\#\#\# .

\end{document}